\documentclass{llncs}
\usepackage{latexsym}
\usepackage{url}

\title{On the defence notion\thanks{This work has been supported by  project ARA Virus.}}

\author{Anne Bonfante\inst{1} \and Jean-Yves Marion\inst{1,2}}
\institute{Loria-INPL \\ \'Equipe Carte \\ \email{anne.bonfante@libertysurf.fr,Jean-Yves.Marion@loria.fr}
           \and
           \'{E}cole Nationale Sup\'{e}rieure \\ Mines de Nancy}

\begin{document}
\maketitle
\section{Computer virology and art of war}
``Trojan horses'', ``logic bombs'', ``armoured viruses'' and 
``cryptovirology'' are terms recalling war gears. In fact, concepts of 
attack and defence drive the world of computer virology, which looks like a 
war universe in an information society. This war has several shapes, from 
invasions of a network by worms, to military and industrial 
espionage{\ldots}

For convenience, and because the term ``virus'' has a bigger impact from an 
epistemological point of view,

(the word ``virus'' was adopted after theoretical works of Cohen \cite{r1} and 
Adleman \cite{r2}) we will refer to ``virus'' or all the different kinds of 
computer infections. The reader may consult Filiol's books \cite{r3,r16} for 
further information about the definitions and classifications of malware.

We could think at first sight that questions about computer security, and 
particularly about fights against computer viruses, are just a matter of 
highly trained security computer officers, and that a victory just depends 
on scientific and technical knowledge. But sometimes algorithmic and 
programming are not enough to catch the whole picture. Indeed, the study of 
malware strategies shows that they are close to the battle of the Horacii 
against the Curiacii{\ldots}

A different point of view could then bring to computer virology a new 
prospect. This study is a part of a general investigation, which tries to 
understand viral strategies in order to anticipate attacks and improve 
defences, and to define the philosophical and political issues. Reading 
famous strategists and philosophers could give precious information about 
conscious- or unconscious viral practices, explicit- and implicit behaviour 
of the malware's writers, whatever kind of attacks they launch. For this, a 
case study is a good way to make the connection between the technical or 
scientific profiles of a virus and the kind of strategy used (the meaning of 
the word "strategy" is here conveyed by political and war philosophies). The 
case of the Whale virus can be extended to the Bradley concept, its 
epistemological model, which is meaningful.

The question of War, as well as the best way to do it and the political- and 
human issues related to it, is very old in the history of philosophy, as we 
can already find it in the \textit{Peloponnesian war }\cite{r4}, written by Thucydide, a Greek general and 
historian in 430 BC, or Plato's \textit{Republic }\cite{r5} (420-340 BC). Philosophy of war has 
always been an active theme of philosophy and was the main issue for some 
authors like Machiavelli (1469-1527), who wrote the \textit{Art of war} \cite{r6} in 1519.

Are the writers of malware Machiavelli's readers? This question sounds like 
a joke, but is legitimate, as we shall see. Indeed, the more dangerous 
viruses are the ones which seem to apply principles and to use strategies of 
the art of war, like Machiavelli's, or from other war theorists like 
Clausewitz \cite{r7}.
\section{Bradley defences}
The Whale virus challenged Scotland Yard for about two weeks during the 
1990's \cite{r8}. This virus is a typical case study, because it has its own 
defensive weapons. From the Whale virus, a virus shape named Bradley was 
proposed by Filiol \cite{r9}. This shape, which could also be considered as an 
epistemological model, will guide us in our study on virus defences. Indeed, 
it uses very complex notions coming from mathematics and computer science in 
order to defend and attack.

Bradley protection systems are first used to prevent the detection of the 
viruses by an antivirus software, and then to bypass the defences of the 
host system, which is under attack. We will study the two most interesting 
types of defence: the armoured code aspect and the virus's one because they 
are typical and relevant. It is worth to notice that the furtive aspect was 
a Whale feature.

An armoured code consists in protecting a code from an analysis, which can 
be static by disassembling, or dynamic, by monitoring executions.

Protecting a code against static analysis is a scientific challenge. The 
paper of Barak {\&} al \cite{r10} on obfuscation is a good illustration of these 
difficulties. With the benefit of hindsight, all the protection methods used 
lean on the same approach. The code protection is performed by an 
obfuscator. In fact, the latter is a kind of compiler, which transforms a 
source programme $s$ into a scrambled programme $s'$ in such a way that:

\begin{enumerate}
\item The scrambled programme $s'$ computes the same thing as the source programme $s$. (in other words, there are semantically equivalent).
\item The runtime of the scrambled programme $s'$ is close to the one of $s$. (that is up to a polynomial).
\item The programme $s'$ is unreadable as well for a human analyst as for a de-obfuscation programme.
\end{enumerate}
The third point corresponds to the obfuscation clause, which turns out to be 
difficult to formalize. Obfuscation consists in rewriting the code in order 
to make it less understandable. Cryptography plays a crucial role in this 
task. The interested reader may consult the recent paper of Beaucamps and 
Filiol \cite{r12} about the practical obfuscation methods. Moreover, a virus may 
mutate when it duplicates with polymorphism or metamorphism techniques. 

An obfuscator protects a programme in a way that we could call passive, 
unlike furtivity techniques. It aims to camouflage a code in order to slow 
the analysis of its adversaries down. Advantage for the attacker with this 
method: the guarantee of immunity for a while, because the time spent to 
camouflage is ever shorter than the one spent by the adversary to detect it. 
Thus, the attacker protects itself against antivirus software, which, on the 
one hand search for a signature of a known malware. On the other hand, it 
protects itself against the work of analysts, which are trying to understand 
the meaning of its code. This method used by virus looks like both a 
camouflage and a shield, which are designed to resist an attack as long as 
possible. It appears then to be a method of passive defence. 

A Bradley virus also combines an active defence method: "furtivity". It 
consists in deceiving a host system, by modifying, for example, system 
interruptions. In the case of Whale, techniques of furtivity allow to detect 
a debugger. The defence becomes active, because a virus can trigger an 
action off, if it is aware that some agent is trying to analyse it. 
Furtivity then allows the prevention of behavioural analysis techniques used 
by antivirus. Moreover, the resistance of a virus to such an analysis allows 
the collection of information about the defence methods of the host system. 
Consequently, we may suppose that this fact is a piece of its attack plan 
and the reasons why the virus was launched{\ldots} We must then look at the 
reasons for which a virus like Bradley integrates so many defence 
techniques.
\section{Why does a virus defend itself?}
The part of the virus code which purpose is to infiltrate a system and to 
deceive the antivirus can be considered as both an offensive weapon (it 
enters a fortress) and a defensive weapon (camouflage). A particularity of 
Whale, as well as the Bradley concept, is that its code contains advanced 
defensive functionalities, which go beyond the traditional ones used to 
enter a system. If a virus was a mere offensive weapon, why allowing such 
defence mechanisms then? 

If we send troops to attack a place, and whatever the issues are, it seems 
normal to arm them and to give them some means to protect themselves. Such 
defensive means are necessary in order to protect troops and avoid 
casualties. However, in the context of a conflict in the information 
society, what does the attacker try to protect?

If the only goal is to get some information back, the loss of a virus is a 
part of the operation. Why are viruses protected then? We think that the 
host systems should address those questions in order to prepare its defence.

This brings us to a first hypothesis. The attacker is a Machiavelli's reader 
without knowing it. Indeed, one of the fundamental principles of the art of 
war is that we must have both offensive and defensive weaponries. A good 
defence is meaningless if it is not armed: the only guarantee of autonomy 
and independence is the ability to be in a defensive position as well as in 
an offensive one if necessary. Computer security questions should then 
necessarily consider the possibility to prepare and use offensive methods. 
However, state legislations, like the French one, do not allow attacks and 
force security officers to set only defensive methods with no possibility of 
counter-attack. The efficiency of the defence is also related to the 
questioning about this ban on possessing and training with its own 
(offensive) weapons. Machiavelli's readers know how important it is to keep 
control over one's own weapons because it is a \textit{sine qua non }condition for security and 
independence. 

The second hypothesis is that a virus provides a defence because it has some 
hostile intentions. If there are as sophisticated high-level defence 
mechanisms as the ones mentioned above, it means that the virus was made for 
a very important purpose. We could use this criterion in order to set a 
typology of viral attacks: the more elaborated the defence is, the more 
important the objective is. 

Lastly, the third hypothesis is that we should call security policy views 
into question. If a virus based on the Bradley concept integrates such 
elaborated sequences of defence, it is because it probably expects a serious 
counter-attack. The defences are thus stronger than what they claim or than 
they are known to be (but it is maybe a trick: fooling the adversary while 
looking more vulnerable than you are in reality.)

Other hypotheses are conceivable, and some show that 'viral war' takes 
sometimes paradoxical or new shapes we should study in order to anticipate 
the future.
\section{New time scales?}
Information technology war implies to have to reconsider the scales of the 
conflicts, particularly the time scale, but also the one of the space. The 
time necessary to decipher and analyse a virus (remember the two weeks which 
were necessary to analyse Whale!) is a ``long'' time. It means that it 
corresponds to a human time-scale. On the other hand, if we consider that a 
viral attack symbolically represents a hostile operation, the length of the 
attack is very short (between a second and a minute). There are at least two 
levels of analysis to bring out in order to understand the defence question:

\begin{itemize}
\item The human scale. The conception of dangerous viruses is a long process because they integrate a lot of mathematics and a big knowledge in computer science. It is hard to capture a virus and difficult to analyse it. This is the case of Whale.
\item The computer scale. A virus is inside a system, and must fight against an anti-virus. The fight lasts a few seconds, or minutes, programme versus programme.
\end{itemize}
We could think at first sight that a viral attack is a completely new kind 
of war, since it takes place in an entirely different way compared to 
traditional war. Can we use the term of war again, since there is nothing in 
common between a few-second strike and a real battle in the field?

A virus writer finds time to design a virus efficiently and cannot ignore 
that it will require a lot of time to decipher and analyse the virus, if it 
is captured. Therefore, this time was anticipated and should be considered 
as a genuine part of the attack. It can be used to create a diversion or to 
have more time (indeed, during the phase of analysis, the attacker can do 
something else). Then this type of virus corresponds more to a traditional 
war, and consequently sets conflicts back in a long time-scale and not in an 
instantaneous one. Isn't it the aim of the attacker? So don't we have to 
consider that a viral attack is just a weapon as another one, or just a 
part, of a global conflict? Such a viral attack is a way to set things back 
in a better-known field, which is more traditional, in a way more human: the 
field of men, logistics and ``normal'' time.

This shift of scale between viral attacks and traditional fights allows us 
to ask further questions. For example, look at the ratio between the 14 days 
of analysis, which is the time necessary to understand Whale, and 1 minute, 
which will be considered like a time-reference at the computer level. The 
ratio between these scales is around a million. The change of scale directly 
implies that when a viral attack is launched, it cannot be controlled at the 
human level anymore, because it is too fast. Admittedly, the use of a gun 
implies that its process is autonomous and does not depend on a human (when 
a gun is fired, the bullet is unavoidable{\ldots}except in "Matrix"!). But 
in the case of a computer war, we should not consider a single virus, but a 
set of viruses against defences. Each of these agents communicates with the 
others, reacts to its environment and takes decisions, independently from 
any human interventions. Moreover, if the computer scale is very short, the 
conflict area could be the whole net \cite{r13}! This change of the scales maybe 
implies that we should change our classical concepts of war analysis, as we 
traditionally find them in books of great authors like Clausewitz \cite{r7}. 
\section{A central question: the losses}
Cyberwar implies to reconsider a crucial notion of every armed conflict: the 
one of the casualties. Indeed, why does someone program an armoured virus? 
What is the attacker's objective? What may he lose, and which losses does he 
want to avoid?

In the setting of traditional war with different armed forces, the question 
of the casualties is one of the most important to make strategic and 
tactical choices. This question is coupled with an obvious moral issue. This 
issue is represented by the assessment of the inherent ratio in any 
operation: the (human-, technical- or practical) cost of an action must be 
related to the interest or to the benefit of this action. So, this assumes 
the responsibility of the military leaders and of the governments, 
particularly a moral responsibility, since casualties should be avoided. 
Taking a risk should then be justified with respect to the current issue, 
especially when human lives are concerned. The moral responsibility is more 
difficult to take, but it just exists in the case of real wars, because we 
can always know the number of deaths, without always really understanding 
the benefit of an operation. Cyberwar changes traditional war categories, 
since the moral and the casualty question as well as the question of the 
just cause, do not come up anymore.

Inversely, the heroic figure or the heroic action supposes that the loss of 
a life is there for a great collective benefit. The greater the sacrifice of 
the hero is, the more he takes risks; the more conscious he is that he will 
not survive to his action, the more heroic he is. We can ask if hackers 
identify to the issues of the hero, as we already mentioned, by overcoming 
all difficulties and responding to all challenges. Indeed, the point for a 
hacker is here to accomplish a technical exploit giving him the feeling of 
being a hero, even if his actions are reprehensible. Recent movies show that 
clearly: it is the heroic figure of Neo in Matrix.

It seems to us that we should understand the specificity of modern wars with 
respect to how they consider casualties. In the case of a virus, the loss of 
the latter is somehow immaterial, maybe anticipated, or even programmed by 
its designer. It means, first, that we do not consider losses as real ones 
(like events to avoid), or at least that it is a part of the strategy, 
especially as we can always replace a virus: losses are duplicable because a 
virus can be duplicated.

Another characteristic of computer-war is that it does not lean on 
economical and industrial resources, which are decisive factors in the case 
of a ``real'' war, even if they are not the only ones. In the case of a 
viral war, the attacker's strength is not directly related to economical- or 
industrial resources, but to a scientific and technical knowledge, which 
then becomes a possible cause of war, or at least reasons for a competition.

When an attacker "loses" a virus, the latter can be analysed. From then on, 
we may reproduce quite easily its design. However, there are some interests 
in avoiding the loss of a viral weapon in order to protect knowledge, in 
terms of cryptography for example. The attacker, as a war leader or a 
strategist, measures the interest of an action and determines the risk he 
runs. One of the clearest specificities of a modern or post-modern conflict 
is that the technical questions aim more and more at replacing the question 
of force. Violent confrontation is shifted to technical considerations.
\section{Counter-attack and issues}
The viral defence system often allows the attacker to stay anonymous, which 
prevents possible counter-attacks. But this fact changes the concept of war 
because the enemy is not identified. What is peculiar to war is to have an 
enemy, a known opponent. Once the enemy remains anonymous, confrontation 
does not correspond to the representation of war, but more to guerrilla 
warfare: the enemy is not conventional and cannot be spotted.

Furthermore, if the attacker tries to hide his strategy as long as possible, 
it means that the host system itself can be protected. Indeed, a virus is 
able to delude a defence in such a way that it can be in position to observe 
tools used for defence: a virus is able to be behind the line, and so is 
able to list all defensive tools. Therefore, this means that the virus code 
has a mechanism the function of which is to detect the tools used by the 
defence in order to analyse intruders, viruses and so on.

We could conclude for the time being that the strategy used by the virus in 
this case is a prospective one. Detecting the defence tools has a meaning 
only if another attack is planned in the near future. Consequently, we 
should analyse the aims of the virus from a temporal point of view. A viral 
attack can be an initial assault before another one, so we should not simply 
consider what represents a given attack at a given time.

Lastly, the issues of computer virology are directly related to other 
domains where computer networks are used as tools to exchange information. 
It is a way to say that the field of viral war is also economic and 
financial, and that it implies political positions, in particular questions 
about individual rights. Economic issues are not always perceptible behind 
some behaviour, which may have some moral justifications: for example, it is 
the case of Sony rootkit \cite{r15}. The reason of the integration of computer 
virology was to protect author rights, and to protect the data contained in 
the disk. The reality was that this technology allowed to spy the user's 
data (it is of great interest from a commercial point of view{\ldots}). The 
ambiguity is in the fact that Sony was perfectly able to justify this 
functionality, which is usually used by malware, in order to protect author 
rights. This worry is legitimate, but nothing proves that the real purpose 
was this one{\ldots} So, computer virology is the place of different kinds 
of wars, but the hostility logic often remains the same.
\section{Evolutions and perspectives}
Viral defence, as a large part of computer security, depends on the results 
of theoretical computer science. For the record, there are Cohen's results 
on viral detection undecidability: it is not possible to construct perfect 
anti-virus software. Other results depend on conjectures of the algorithmic 
complexity theory (like if P=PSPACE, then practically all cryptographic 
codes will be broken). It is interesting to see once more that we face a 
scale problem. Indeed, conjectures related to the algorithmic complexity 
theory imply that a given problem is computable but the runtime is beyond 
belief and is greater than the age of the universe! Science acts as a 
guarantor for defence soundness. If a defence depends on the algorithmic 
complexity theory, so it must take into account the time scales that we have 
mentioned previously. A viral defence may be conceived to resist an analysis 
at a human scale, and from then on can be considered as immune with respect 
to the computer time scale (where the time scale is about a minute). On the 
other hand, during a computer attack, and at a computer time scale, this 
heavy defence slows the virus execution down. So, the issue is to choose the 
right weaponry according to the targets that are aimed. This subject was 
widely discussed in many strategy books, as Machiavelli did in his \textit{Art of war}.

As we saw it above, viral war integrates the necessity of a prospective 
strategy: on the one hand, because it could use the information the virus 
found, pursuing the elaboration of its attack's tactics according to the 
environment it is located in. In this sense, a virus is a modern weapon 
since it is adaptable: there are interactions between a virus and its 
environment. There is an epistemological and scientific dimension as soon as 
we think about a viral attack as a deployment of a large number of 
autonomous and cooperative agents. This direction recalls analogies with 
biological immune systems \cite{r14}. Another possibility is to consider the 
agents of a network with divergent interests like in the algorithmic game 
theory. The game theory, which tries to predict behaviours of a population 
in a given environment, could then be used to analyse political and 
sociological behaviours of agents running viral attacks, and also of the 
users who are concerned by those attacks.

To come back to the question of the interaction of a virus in its 
environment, we can say it is inseparable from a forecast. That is that a 
virus adapts to its environment in order to pursue an attack, or to send 
information back with the view to make another attack. This temporal 
forecast, even if it is implicit, is the sign that a viral arm cannot be 
understood, or analysed, without thinking of the one who possesses it. The 
attacker's intention (or the intention of the ones who order this action) is 
revealed by the question, which seems harmless at first glance: what is the 
purpose of a weapon? If, to be specific, it can be used to make another 
attack, or just to pursue the current one: this attack is a step. It also 
means that it is necessary for the ones who are fighting against viral 
attacks, not only to view an attack \textit{here and now}, but always to think about the 
intentions that could be implied. To give another conclusion to this 
discussion, we should analyse a viral attack by what it implicitly 
discloses: the virus's method for the attack indicates what its writer 
knows, or what he thinks he knows about the target he aims. Anybody thinking 
he is a potential target should integrate into its security policy what he 
wants to show to the others, in such a way that he directs the kind of 
attack he can be the target of. The issue at stake here is to apply the good 
old strategy: a false lure, a false target, or a noticeable weakness, which 
transforms stratagems into an art of playing with representations and to 
trick enemies. We must always come back to this Machiavelli's principle: 
``\textit{Never believe that the enemy doesn't know what he does}'', always assume that you have to deal with a smart-, surely crafty 
enemy, whose intentions are not always visible. So, the security of a target 
should not only stem from the strength of a defence: the attackers will 
always succeed to penetrate by a way or another. We should also know how to 
use trickery, how to play with psychological effects and make our opponent 
believe that defences are different from what they really are{\ldots}
\section*{Acknowledgement}
We would like to thank first lieutenant de Marqueissac for her valuable help 
in improving the correctness of our English.

\end{document}